\documentclass[a4paper]{article}

\usepackage{INTERSPEECH2022}
\usepackage{multicol}
\usepackage{multirow}
\usepackage{wrapfig}
\usepackage{subcaption}
\usepackage{todonotes}
\usepackage{booktabs}
\usepackage{caption}
\usepackage{hyperref}
\usepackage{comment}
\title{Mel Frequency Spectral Domain Defenses against Adversarial Attacks on Speech Recognition Systems}
\name{Nicholas Mehlman, Anirudh Sreeram, Raghuveer Peri, Shrikanth Narayanan}

\address{
  USC Ming Hsieh Department of Electrical and Computer Engineering, Los Angeles, CA 90089}
\email{\{nmehlman, asreeram, rperi, shri\}@usc.edu}

\begin{document}
\captionsetup{width=0.5\textwidth}
\maketitle
\begin{abstract}

A variety of recent works have looked into defenses for deep neural networks against adversarial attacks particularly within the image processing domain. %However, automatic speech recognition (ASR) has become increasingly ubiquitous in daily life.
Speech processing applications such as automatic speech recognition (ASR) are increasingly relying on deep learning models, and so are also prone to adversarial attacks.
However, many of the defenses explored for ASR simply adapt the image-domain defenses, which may not provide optimal robustness. This paper explores speech-specific defenses using the mel spectral domain, and introduces a novel defense method called `mel domain noise flooding' (MDNF). MDNF applies additive noise to the mel spectrogram of a speech utterance prior to re-synthesising the audio signal. We test the defenses against strong white-box adversarial attacks such as projected gradient descent~(PGD) and Carlini-Wagner~(CW) attacks, and show better robustness compared to a randomized smoothing baseline across strong threat models.

\end{abstract}


\section{Introduction}

Recently there has been growing interest in defending automatic speech recognition (ASR) systems against adversarial attacks. In an adversarial attack, a malicious actor attempts to cause the ASR model to miss-transcribe an utterance by introducing small amounts of carefully crafted perturbation to the \textit{benign} audio. 
%Common attack algorithms, such as the fast gradient sign method (FGSM) use the ASR model's gradients in a gradient ascent procedure that attempts to maximize a loss function between the benign and adversarial transcriptions \cite{explaining_adv_examples}.
%A majority of defenses have been proposed largely inspired by similar work performed in the image domain. 
A vast majority of the existing defenses against such attacks are inspired by similar work in the computer vision domain. For example, randomized smoothing was originally proposed for images by~\cite{certified_smoothing}, but has been applied to ASR (e.g. \cite{adversarial_attacks+defenses}). 

While these methods often provide acceptable adversarial robustness in the audio domain, they do not take full advantage of the unique properties of speech signals. Speech is highly structured in both time and frequency in a manner that distinguishes it from images. Therefore, defenses that have been shown to be effective in the image domain (e.g., randomized smoothing) may not be optimal for ASR applications.
For example, the authors in \cite{wavegard} have demonstrated the enhanced robustness of techniques that leverage the mel spectrogram and linear-predictive coefficient (LPC) representations.
% Indeed, previous work has demonstrated the merits of using speech-specific defenses. 
% For example \cite{wavegard} shows success with using the mel spectrogram and linear-predictive coefficient (LPC) representation domains. 
Along these lines, this paper explores novel ways to leverage the mel domain in a defensive framework.

We find that the process of transforming the speech signal into the mel representation, and subsequently re-synthesising the audio can provide a defense against adversarial attacks. Additionally we introduce a novel \textit{noise flooding} technique termed as `mel domain noise flooding' (MDNF) that injects white Gaussian noise~(WGN) into the mel representation prior to re-synthesis\footnote{We adopt the term `noise flooding' from \cite{noise_flooding}}. 
Robustness is further enhanced by shaping the variance of the noise to match the frequency distribution of adversarial perturbations. We show that MDNF substantially improves the adversarial performance of a relatively weak defense based on mel re-synthesis, and retains the robustness in the case of an already well-performing re-synthesis model. We also show that our proposed defense outperforms randomized smoothing across a range of attack threat models. 
%(a competitive baseline used in prior works \cite{certified_smoothing} \cite{adversarial_attacks+defenses}) across multiple threat models.

%In this paper we introduce mel-domain noise flooding (MDNF) a pre-processing defense that utilizes structural properties specific to speech signals, and thus is well suited for use in ASR systems. Additive white Gaussian noise is introduced to the two-dimensional mel representation of the speech signal, and the resulting spectrogram is inverted to produce a time-domain signal. The inversion process uses a Generative Adversarial Network (GAN), and we
%we demonstrate the effectiveness of our approach using waveGAN \cite{wavegan} and melGAN \cite{melgan} architectures. 
%The proposed defense outperforms randomized smoothing (a competitive baseline used in prior works \cite{certified_smoothing} \cite{adversarial_attacks+defenses}) across multiple threat models.

\section{Background}

\subsection{Adversarial attacks}
Adversarial attacks can be broadly separated into two categories based on their level of knowledge of the ASR model's internal state \cite{attacks_defense_in_DL}. Black-box attacks, as the name implies, are completely agnostic to the inner workings of the model. White-box attacks have complete knowledge of the ASR, and in particular can extract loss gradients (with respect to the input audio) to construct perturbations. Adversarial attacks can also be classified based on their objective \cite{attacks_defense_in_DL}. Untargeted attacks simply attempt to `fool' the ASR system (i.e. cause it to produce an incorrect transcription), while targeted attacks try to produce a specific erroneous output. For a more detailed description of the different attacks, we refer the reader to \cite{attacks_defense_in_DL}. 
%(see for example \cite{attacks_defense_in_DL}). 

One popular white-box attack is the fast gradient sign method (FGSM) first introduced by Goodfellow \textit{et al.} in \cite{explaining_adv_examples}. Here the adversary adds a one-shot perturbation constructed using the sign of the model's gradients. A more sophisticated approach is the projected gradient descent (PGD) attack from \cite{pgd}. Similar to FGSM, the perturbations are constructed using the sign of the gradients, but instead of a one-step process, the PGD attack is generated iteratively. 
As shown in eq. \eqref{pgd_eq}, at iteration $k$ the perturbed signal $x_k$ is projected onto an $\epsilon$-ball (typically using the $\ell_2$ or $\ell_{\infty}$ norm) around the original input to enforce an adversarial perturbation that is minimally noticeable \cite{pgd, adversarial_attacks+defenses}.
\begin{equation}
 x_{k+1} = \mathrm{Proj}_{\epsilon} (x_k + \epsilon_{step}( \nabla_{x_k}(\mathcal{L}(\phi(x_k),y_b)) )) 
\label{pgd_eq}
\end{equation}
Here $\mathrm{Proj}_{\epsilon}$ represents projection onto the $\epsilon$-ball, and $\mathcal{L}(\phi(x_k),y_b)$ is the loss function between the ASR transcription of the perturbed signal $x_k$ and the benign transcription $y_b$. $\epsilon$ is a hyper-parameter that controls the relative strength of the attack~\cite{pgd}. An alternative version of PGD uses a minimum signal-to-noise ratio (SNR) as a bound for the perturbation magnitude. Table \ref{tab:pgd_params} provides a brief description of the other PGD parameters. 

Another common attack is the targeted Carlini-Wagner (CW) attack. The CW attack is based on a constrained optimization problem that searches for a perturbation $\delta$ with \mbox{$\| \delta \| < \tau$} which produces the target miss-transcription \cite{carlini-wargner}. By repeatedly solving this problem with gradually smaller values of $\tau$, the attack enforces a minimally-perceptible modification that still obtains the desired result.

\begin{table}[tb]
    \centering
    \caption{PGD attack hyper-parameters corresponding to eq. \ref{pgd_eq}. The perturbation magnitude for untargeted attacks is controlled by $\epsilon$, while targeted attacks are characterized by minimum SNR.}
    \label{tab:pgd_params}
    \resizebox{0.48\textwidth}{!}{
    \begin{tabular}{|c|c|}
         \hline
         \textbf{Parameter} & \textbf{Description} \\ 
         \hline
         $\epsilon$ or SNR & maximum allowed perturbation \\ 
         \hline
         norm & norm for constraining perturbations (i.e. $\ell_2$, $\ell_{\infty}$) \\ 
         \hline
         max iter. & max number of iterations used to construct attack \\
         \hline
          $\epsilon_{step}$ & learning rate used for constructing attack \\ 
         \hline
    \end{tabular}
    }
\end{table}

\subsection{Defenses}

One common approach to making a neural network more robust to adversarial attacks is adversarial training \cite{intriguing}, or integrating adversarial examples as a part of the model's training data. %This is described by Szegedy \textit{et. al.} in \cite{intriguing}. 
%This method builds off of other commonly-employed methods for data augmentation, such as adding Gaussian noise to an audio signal, or applying random rotations to an image. 
%In \cite{asr_training_augmentation}, Sun \textit{et al.} use FSGM to construct `real-time' adversarial examples for each training mini-batch, and report some degree of adversarial robustness in the ASR scenario. 
This approach has been used to develop robust ASR systems \cite{asr_training_augmentation}.
%Since these examples use the current state of the model parameters to generate perturbations, they are more informative than static adversarial samples created offline prior to training \cite{asr_training_augmentation}. 
%Using the Aurora-4 corpus, they report a $14\%$ reduction in word-error-rate~(WER) over models trained on benign samples alone. 
While adversarial training may provide improved robustness, it generally results in degraded performance on benign samples~\cite{jati2021adversarial}. Additionally, since it is implemented during ASR training, it cannot be deployed with pre-existing ASR systems.

%In \cite{explaining_adv_examples}, Goodfellow \textit{et. al.} present an alternative approach which modifies the loss function as opposed to the data-set. In particular, they add an additional term that penalizes rapid changes in the prediction manifold that might be leveraged by an attacker \cite{explaining_adv_examples}. This results in a reported $10\%$ reduction in WER over the original model \cite{explaining_adv_examples}. Similar to adversarial training, this requires re-training the model, which can be computationally infeasible.
Another popular defense (with certifiable guarantees) is randomized smoothing~(RS), which involves the introduction of WGN to the input of the ASR system \cite{adversarial_attacks+defenses, certified_smoothing}. Predictions are made by averaging the model's logits over several different random perturbations. Provided the ASR model is robust to additive noise, introducing these stochastic perturbations to the input can counteract the carefully crafted adversarial perturbations constructed by the attacker \cite{adversarial_attacks+defenses, certified_smoothing}. 
%Cohen \textit{et al.} in \cite{certified_smoothing} prove that this procedure provides certifiable robustness to attacks constrained in the $\ell_2$ norm. 
Despite the good performance, RS increases computation time during inference due to the need for multiple forward passes per sample.

A variety of more sophisticated defenses utilize re-synthesis of the speech signal as a way to `discard' adversarial content. In this category of defenses, recently developed systems use generative adversarial networks (GANs) to produce a facsimile of the original input signal that retains the primary content of the speech \cite{wavegan}. GANs consist of a generator, which seeks to produce realistic data, and a discriminator that attempts to distinguish legitimate data samples from the synthetic samples produced by the generator. 
%Through adversarial training of the generator/discriminator pair, the former can learn to produce highly plausible outputs. Esmaeilpour \textit{et al.} introduce a `class conditional GAN' defense in \cite{classconditionalGAN} which transforms a random-noise vector into a speech-like output, conditioned on the discrete wavelet representation of the original signal. Their defense operates in an iterative manner that searches for the noise vector that minimizes the chordal distance between the original and synthesized speech \cite{classconditionalGAN}. However, this optimization procedure incurs a significant computational cost. 
In \cite{adversarial_attacks+defenses}, Zelasko \textit{et al.} use the mel domain in conjunction with the generator portion of the WaveGAN vocoder introduced in \cite{wavegan}. It takes the mel spectrogram as a conditioning input, and attempts to re-synthesize the audio in a single pass. They demonstrate that introducing the vocoder as a pre-processor to the ASR system improves the model's performance under the PGD, FSGM, and imperceptible attacks \cite{adversarial_attacks+defenses}. 
%Prior work (e.g. \cite{wavegard}) has also demonstrated the defensive capabilities of the mel domain.

\subsection{Mel Domain Representation of Speech Signals}
\label{sec:mel_rep}

The mel domain represents a speech signal in a time-frequency format specific to the structure of auditory processing and human speech. 
Mel spectrograms are commonly used as features for speech recognition and other speech applications \cite{davis1980comparison}. The conversion to the mel domain is lossy, and significantly reduces dimensionality as compared to a raw audio waveform. However, it still retains the core information relevant to speech related tasks \cite{rabiner}.

The time domain speech signal is converted to the mel representation by the following procedure. First, the vector of speech samples is segmented into overlapping fixed-length \textit{frames}. 
%$x^{(m)}[l], l = 1,..,N$ where $m$ is the frame index and $N$ is the frame length \cite{rabiner}. 
Such frame-level processing leverages the stationarity of speech signals -- the statistics remain constant over short time intervals. Next, the discrete Fourier transform (DFT) of each frame is computed. %to generate the frequency domain representation $X^{(m)}[k]$. 
Finally a mel filter-bank consisting of triangular filters is applied. 
%$V_r[k], r = 1,...R$ where $R$ is the number of filters \cite{rabiner}. 
The frequency response of the mel filter bank is inspired by the `critical bands' of human hearing \cite{rabiner}. Critical bands model the finite frequency resolution of the auditory system by a series of variable-width band-pass filters. Bandwidth increases for high frequencies where the human ear has been shown to have a coarser resolution \cite{rabiner}.

%%%%%%%%%%%%
\section{Proposed Defenses}
\begin{figure*}
\captionsetup{width=\textwidth}
    \centering
    \includegraphics[width=15 cm]{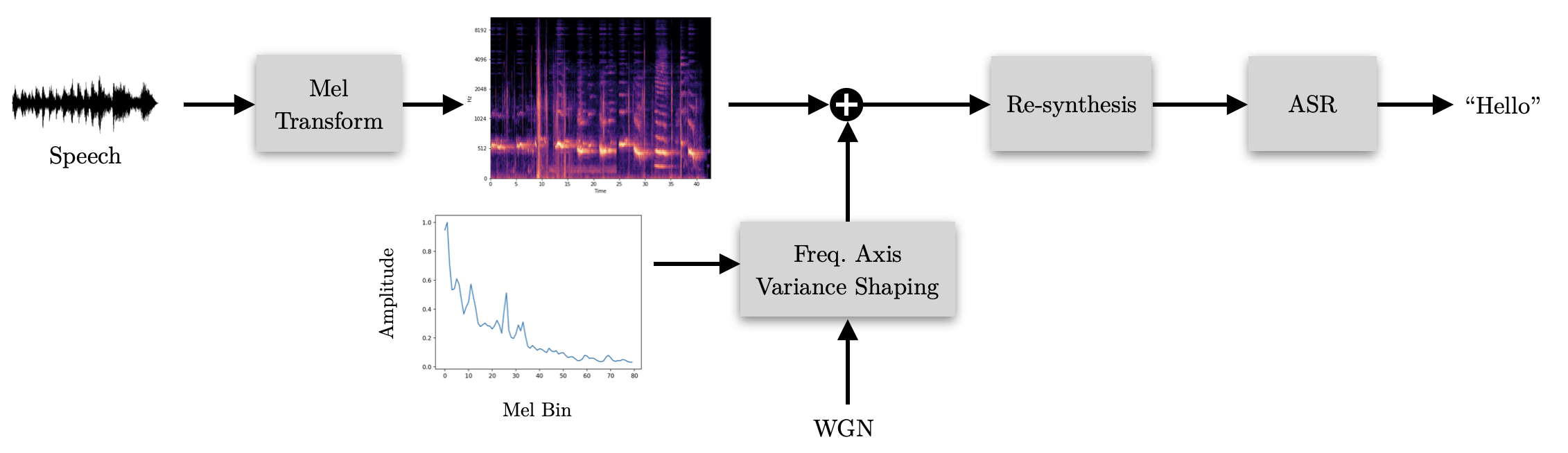}
    \caption{Block diagram of the mel domain noise flooding (MDNF) defense. Additive noise is introduced to the mel spectrogram. The shown curve is an empirical-derived distribution of adversarial perturbations across the mel bins. Noise-variance shaping applies this curve to the noise across the frequency axis for each time-step.}
    \label{fig:block_diagram}
\end{figure*}

We introduce two defenses that utilize the mel frequency representation of speech signals. First we present the mel domain transform as a standalone defense, and then we show how stochastic noise flooding of the mel coefficients can be used to further improve adversarial robustness.

As shown by Hussain \textit{et al.} in \cite{wavegard}, the process of passing a speech signal through the lossy mel transform can itself provide adversarial robustness. First, the time domain speech signal is broken into frames and passed through a mel filter-bank to produce a mel spectrogram. Since the mel representation of a speech signal has limited time and frequency resolution, the process of converting a speech signal to a mel spectrogram is inherently lossy. However, given the domain's speech-specific design discussed in the previous section, it is reasonable to expect that the information relevant to the linguistic content of the utterance will be retained. The same cannot be assumed for the non speech-like adversarial perturbations. Therefore the transform is more likely to preserve speech content than it is to retain information added by the adversary \cite{wavegard}. In order to re-synthesize the time-domain speech, we use two different GAN-based generators: the `waveGAN' \cite{wavegan} which has been explored as a defense in \cite{adversarial_attacks+defenses} and the  `melGAN' presented in \cite{melgan}. Both models are similar, although unlike the waveGAN, the melGAN does not require a random noise vector as input in addition to the mel spectrogram. 

%The motivation for leveraging the mel domain in our defenses follows from the work presented by Hussain \textit{et al.} in \cite{wavegard}. Since the mel representation of a speech signal has limited time and frequency resolution, the process of converting a speech signal to a mel spectrogram is inherently lossy \cite{wavegard}. However, given the domain's speech-specific design discussed in the previous section, it is reasonable to expect that the mel representation will retain the information relevant to the linguistic content of the utterance \cite{wavegard}. However there is no reason to believe that the information contained within the perturbations will be preserved during the mel conversion. In other words, the transform is much more likely to preserve speech content than it is to retain information added by the adversary \cite{wavegard}.
%Hussain \textit{et al.} support this hypothesis by showing that overcoming a perceptually motivated defense requires adversarial perturbations approximately $6$ times larger than those require to break a similar non-perceptual defense \cite{wavegard}.

To further improve the performance of the vanilla mel domain defense, we modified the above procedure to introduce mel domain noise flooding. As shown in fig. \ref{fig:block_diagram}, the conversion from time to mel domain representation is performed using the standard procedure described in Section~\ref{sec:mel_rep}. However, prior to passing the spectrogram to the GAN for re-synthesis we add WGN to the mel coefficients, thereby introducing stochastic perturbations to the mel representation. The variance of the additive noise is a hyper-parameter that determines a trade off between defense robustness and benign performance. 
%Higher noise energy presents a more difficult challenge to an attacking adversary, but also degrades the quality of the GAN's reconstruction.

The most straightforward implementation would be to add WGN with equal variance across all mel frequency bins. However, an empirical analysis of the difference between adversarial and benign samples (produced using an untargeted PGD attack) reveals that the perturbations introduced are not uniformly distributed across the mel frequency bins. Instead they appear to skew disproportionately towards lower frequencies.
%We conjecture that the non-uniform frequency distribution of the adversarial perturbations is reflective of the sensitivity of the ASR: ASR models are known to rely more on the lower frequencies of the signal \cite{}. 
%if the model weights lower mel bins more strongly in the transcription process, then 
%Therefore the adversary likely crafts perturbations that also utilize these same lower frequency regions. 
This finding inspired us to shape the additive noise along the frequency axis such that the variance in a given mel bin is directly proportional to the relative strength of the attack as reflected by the curve shown in Fig.~\ref{fig:block_diagram}. As shown in Fig.~\ref{fig:block_diagram} the noise is shaped prior to being added to the clean mel spectrogram.

%%%%%%%%%%%%
\section{Experiments}

Our defenses were tested within the Armory\footnote{\url{https://github.com/twosixlabs/armory}}
framework using the DeepSpeech2 ASR model \cite{deepspeech}. 
%We evaluated against both targeted and untargeted attacks for a wide range of threat models. 
We compared the performance of our proposed defenses with a competitive baseline, randomized smoothing \cite{certified_smoothing}, across a range of targeted and untargeted threat models. 
%Like the proposed defense, randomized smoothing is implemented as a pre-processor to the ASR model. Additionally, it also utilizes stochasticity in the form of additive noise. 
Our implementation of RS uses additive WGN at $15$ dB SNR and averages the character probabilities over $5$ forward passes.
%TO-DO
%Additionally, two alternative defenses are used as baselines for performance evaluation. 
%We also implemented a reference defense that used a short-time Fourier transform (STFT) instead of the mel domain. This allowed us to see how noise-flooding performed when applied to a non-perceptual frequency domain. 

Based on empirical observations, we found that the defense performed sub-optimally even under benign conditions when used with the pre-trained DeepSpeech2 model. This is likely because the re-synthesized signals generated by the waveGAN and melGAN have audible artifacts, creating a mismatch with the signal expected by the ASR model. 
%In other words, they exhibit substantial perceptual differences from the `clean' samples used to train the original ASR. 
To address this discrepancy we fine-tuned the DeepSpeech2 ASR on the re-synthesized samples (this was performed separately for the waveGAN and melGAN producing two fine-tuned ASRs). Based on preliminary experiments, we found that a mixture of $70\%$ re-synthesized and $30\%$ clean samples with white Gaussian noise added to all samples produced a model with an acceptable benign ASR performance close to the original ASR model without re-synthesis.

The shaping curve shown in fig. \ref{fig:block_diagram} was produced empirically using $100$ pairs of adversarial and benign samples. By considering the difference between these pairs, we were able to isolate the adversarial perturbations and thus determine their average mel-domain distribution. 

\textbf{Dataset:}
Our experiments used the Librispeech dataset which consists of approximately $1000$ hours of speech utterances sourced from audio books \cite{librispeech}. For fine-tuning of the ASR model the `train clean 100' split was used. As described previously, we used a mixture of $70\%$ resynthesized and $30\%$ clean utterances, with WGN added to all samples.
%After some experimentation, it was found that a mixture of $70\%$ melGAN samples, $30\%$ clean samples all with additive white-Gaussian noise produced the best-performing model. 
We performed initial evaluations and tuning of the noise-flooding amplitude on the first $500$ samples of the `dev clean' set, and used the first $500$ samples of the `test clean' split in the final evaluation.

\subsection{Adversarial attacks (Threat models)}
We test four different defenses: melGAN only (MG), waveGAN only (WG), melGAN with MDNF (MG-MDNF), and waveGAN with MNDF (WG-MDNF) in addition to the RS baseline. Defenses were evaluated against both targeted and untargeted PGD attacks. We performed experiments with various norms and $\epsilon$/SNR values (as described in Table~\ref{tab:pgd_params}) to investigate the robustness of our methods to different conditions. All PGD attacks used $100$ max iterations. To distinguish between the different untargeted PGD scenarios, we will use the notation `norm/$epsilon$'. Targeted SNR-bounded PGD scenarios will be referred to by the SNR in dB. In addition, we also evaluate the defenses against the CW attack\footnote{\url{ https://github.com/Trusted-AI/adversarial-robustness-toolbox/blob/main/art/attacks/evasion/adversarial_asr.py}} with initial $\epsilon=0.01$ and $400$ iterations with learning rate 1e-4. 
We evaluate the defenses using word error rate (WER) between the ground-truth transcription and the transcription produced by the ASR in each scenario. 
Targeted attacks are additionally evaluated by their `target word error rate' which reflects the effectiveness of the adversary in forcing the desired transcription. High values indicate that the adversary is failing to achieve its objective. 

%\todo[]{rperi: Add details of CW attack parameters}

Since MDNF is a stochastic method, we implemented an adaptive attack called expectation over transformation (EOT) with the PGD attack in which the adversary averages ASR gradients across multiple inference calls \cite{EOT}. EOT allows the attacker to estimate the expected value of the gradients under the random perturbations introduced by the defense, and has been shown to produce robust adversarial examples even in the presence of random transforms.

\section{Results}

Table~\ref{tab:results_untargeted} shows the results for the melGAN-based defenses against various untargeted PGD attacks, as well as benign performance (denoted by `Ben.'), which considers the case where the ASR system is not under attack. The MelGAN alone (MG) adds some degree of robustness over the undefended model (defense is None), improving WER on the $\ell_2$/$1.5$ attack by about $3.5\%$ (see rows 1 and 3). However, the addition of noise flooding (i.e. MG-MDNF, row 4) improves performance substantially, reducing the WER by almost $32\%$ over MG. MG-MDNF also outperforms the RS baseline in the stronger (i.e., larger $\epsilon$) attack scenarios. For the `weaker' attack scenario ($\epsilon=0.5$) MG-MDNF produces slightly worse adversarial performance compared to RS (WER goes up by $1\%$). For these experiments the adversarial perturbations are substantially smaller in magnitude than the noise added during randomized smoothing. 
%For example, the adversarial SNR (i.e., the relative amplitude of the adversarial perturbations) for the PGD $\ell_2$/$0.5$ attack is about $30.6$ dB, whereas RS adds noise at $15$ dB SNR. In the $\ell_2$/$1.5$ scenario, the adversarial SNR is only about $21$ dB. 
We hypothesize that RS is better able to overwhelm the adversarial perturbations for the high SNR~(small $\epsilon$) attacks, and thus performs slightly better relative to MDNF. 

We observe a similar trend in the $\ell_\infty$-norm PGD attack scenarios, where the MG-MDNF method outperforms the RS baseline by a large margin in the high $\epsilon$ scenario, while still retaining competitive performance in the low $\epsilon$ threat model.
These results suggest that the proposed MG-MDNF defense shows strong robustness, particularly against high $\epsilon$ attacks. This is despite the fact the MG alone produces only a small improvement in WER over the no-defense scenario, implying that the bulk of the performance gains derive from the noise flooding. This comes at the cost of a small increase in benign WER of about $5\%$ (column 1) compared to the undefended scenario.

We also tested the robustness of the MG-MDNF defense to an adversary using expectation over transformation (EOT). We found that by averaging loss gradients over $5$ calls the adversarial WER on the untargeted PGD $\ell_2$/$1.5$ increased from $52.4$ to $63.3$. A similar effect was observed for the $\ell_\infty$/$0.01$ attack. Therefore MG-MDNF still provides a fairly strong defense even against an EOT adversary. 
%It should also be noted that using EOT is also computationally expensive for the adversary, due to the need for multiple ASR calls per iteration. 

\begin{table}[tb]
    \centering
    \caption{\% WER~(lower is better) for \textbf{melGAN defenses against untargeted PGD attacks}. 
    %MG-MDNF substantially outperforms MG across all threat models. %It exceeds or matches the performance of the randomized smoothing baseline. 
    Results use $500$ samples from the test clean split of the Librispeech dataset. Each column shows the results for a different threat model characterized by $\epsilon$.}
    %All scenarios use a maximum iterations value of $100$.}
    %The $\epsilon$ for each of the threat models is shown below the attack.}
    %Attack threat models are shown as $\epsilon$.}
    \label{tab:results_untargeted}
    % \begin{tabular}{|p{1.8cm}|p{.8cm}|p{1.5cm}|p{1.5cm}|p{1.5cm}|p{1.5cm}|p{1.8cm}|}
    % \hline
    %  & Ben. & $\ell_2$/$1.5$/$100$ & $\ell_2$/$1.0$/$100$& $\ell_2$/$0.5$/$100$ & $\ell_{\infty}$/$0.01$/$100$ & $\ell_{\infty}$/$0.005$/$100$ \\
    % \hline
    % \textbf{None}     & $11.03$ & $87.92$ & $76.00$ & $57.13$ &  $115.42$ & $96.32$ \\
    % \hline
    % \textbf{MG}       & $14.76$ & $84.16$ & $75.12$ & $63.03$ &  $104.00$ & $89.88$ \\
    % \hline
    % \textbf{MG-MGNF}  & $16.00$ & $52.35$ & $42.20$ & $31.01$ &  $65.59$  & $47.08$ \\
    % \hline
    % \textbf{RS}       & $15.69$ & $70.59$ & $51.79$ & $29.96$ & $93.26$  & $49.18$ \\
    % \hline
    % \end{tabular}
\resizebox{0.48\textwidth}{!}{
\begin{tabular}{c|c|ccc|cc}
\toprule
\multirow{2}{*}{Defense} & \multirow{2}{*}{Ben.} & \multicolumn{3}{c|}{PGD ($\ell_2$ norm)}    & \multicolumn{2}{c}{PGD ($\ell_\infty$ norm)} \\ %\cline{3-7} 
                  &                         & 1.5 & 1.0 & 0.5 & 0.01    & 0.005    \\ \midrule
None              & 11.0                   & 87.9   & 76.0   & 57.1   & 115.4      & 96.3        \\ 
RS                & 15.7                   & 70.6   & 51.8   & 30.0  & 93.3       & 49.2        \\
MG                & 14.8                   & 84.2   & 75.1   & 63.0   & 104.0      & 89.9        \\
MG-MDNF           & 16.0                   & 52.4   & 42.2   & 31.0   & 65.6       & 47.1        \\ 
 \bottomrule
\end{tabular}
}
\end{table}

The results for targeted PGD and CW attacks are shown in Table \ref{tab:results_targeted}. In these experiments the MG defense produces a greater improvement in WER over the undefended model than it does for the untargeted attacks (Table \ref{tab:results_untargeted}). For example, the WER for the $20$ dB PGD attack is reduced by around $14.5\%$ over the undefended model (see rows 1 and 3). Once again, however, MG-MDNF (row 4) does markedly better than either the undefended or MG scenarios, further reducing the undefended WER by $48.5\%$. As with the untargeted attacks, MG-MDNF performs better than the RS baseline for higher $\epsilon$ (stronger) attacks, and about the same for the smaller $\epsilon$ scenarios. Additionally, MG-MDNF provides a strong defense against the CW attack, implying that it generalizes well to a range of threat models. 

Table \ref{tab:results_wavegan} shows the performance of the WG and WG-MDNF defenses against a variety of untargeted PGD attacks. Unlike the melGAN, the waveGAN appears to provide a robust defense even without noise-flooding. For the $\ell_2$/$1.5$ threat model, the adversarial WER with the waveGAN alone is $39.2\%$ (row 2). However, the benign performance of the WG defense is slightly worse than that of the MG defense. The addition of noise flooding (WG-MDNF) produces minimal change in adversarial performance, with the a minor ($1.6\%$) increase in benign WER (column 1).  This may be because the waveGAN takes a global noise vector (in addition to the mel spectrogram) as input. Therefore the gradients seen by the adversary are slightly different on each iteration, leading to inherent stochasticity even without noise-flooding. In summary, we find that noise-flooding substantially improves the adversarial robustness of a relatively poor-performing mel-transform defense (i.e., the melGAN). When applied to an already strong defense such as the waveGAN, it largely retains the benign and adversarial performance of the WG defense.
%does not substantially degrade the benign or adversarial performance of the WG defense. 

%We also ran experiments using a short-time Fourier transform (STFT) in place of a mel domain transform. We found that the STFT alone provided essentially no adversarial robustness, but with noise flooding on the frequency domain coefficients the adversarial WER improved by $26.26\%$. This supports the hypothesis that the additive noise in the transform-domain can act to boost the robustness of a weak defense.

As discussed previously, we found that adversarial performance improved when the noise was shaped along the frequency axis in proportion to the empirical distribution of the adversarial perturbations. Without this noise shaping (i.e. with equal variance noise added to each bin) the MG-MDNF defense produced an adversarial WER of $64.0\%$ compared to $52.4\%$ when using the curve shown in Figure \ref{fig:block_diagram} ($\ell_2$/$1.5$ threat model). This particular curve was generated using an $\ell_2$ PGD attack, and seems to perform well across various threat models, including the CW attack. However our experiments suggest that this shaping curve may not be globally optimal: for the $\ell_\infty$/$0.01$ attack, we observed that the unshaped WGN (equal variance) with melGAN significantly outperforms the shaped noise ($25.8$ vs. $65.6$ adversarial WER). 

%We also tested the effect of increasing the maximum iterations (max iter. in Table \ref{tab:pgd_params}) of the PGD attack. We found that increasing from $100$ to $250$ iterations increased adversarial WER in the $\ell_2$/$1.5$ scenario by around $5.5\%$, but that further increasing to $500$ iterations had only a minor impact. This suggests that the number of iterations is not a substantive limiting factor in the adversary's success.

We found that increasing the number of attack iterations (e.g, max iterations = 250, 500) produced only a minor change suggesting that this is not a substantive limiting factor to the adversary's success.

\begin{table}[tb]
    \centering
    \caption{\% WER~(lower is better) for \textbf{melGAN defenses against targeted PGD and CW attacks}. 
    %MG-MDNF outperforms MG, and also matches or exceeds the performance of the randomized smoothing baseline. 
    %PGD results use $100$ max iterations, and 
    PGD results use $500$ samples from the test clean split of the Librispeech dataset. 
    %CW results are for $400$ max iterations, and 
    CW results use $246$ samples from the test clean split. The numbers in parentheses for CW results show the Target \%WER (Higher is better robustness).} 
    %The threat models are characterized by the SNR(dB) for PGD and $\epsilon$ for the CW attack}
    \label{tab:results_targeted}
    % \begin{tabular}{|p{1.8cm}|p{2.2cm}|p{2.2cm}|p{2.2cm}|p{2.2cm}|}
    % \hline
    %             & PGD/$20$dB/$100$     & PGD/$30$dB/$100$   & PGD/$40$dB/$100$   & Carlini-Wagner*  \\
    % \hline
    % \textbf{None}    & $102.54$ ($95.90$) & $91.44$ ($98.41$)  & $58.94$ ($101.04$) &  $99.52$ ($65.25$)    \\
    % \hline
    % \textbf{MG}      & $88.92$ ($101.20$) & $74.06$ ($100.56$) & $50.50$ ($100.83$) &  $21.63$ ($100.57$)  \\
    % \hline
    % \textbf{MG-MDNF} & $54.02$ ($101.41$) & $29.20$ ($101.18$) & $19.92$ ($101.48$) &  $19.25$ ($101.03$)   \\
    % \hline
    % \textbf{RS}      & $72.85$ ($101.01$) & $24.90$ ($101.06$) & $18.05$ ($101.28$) &     \\
    % \hline
    % \end{tabular}
\resizebox{0.48\textwidth}{!}{
%     \begin{tabular}{c|ccc|c}
% \toprule
% \multirow{2}{*}{Defense} & \multicolumn{3}{c|}{Targeted PGD}                         & Targeted CW             \\ %\cline{2-5} 
%                   & $20dB/100$       & $30dB/100$       & $40dB/100$       & $0.01/100$       \\ \midrule
% None              & 102.54 (95.90) & 91.44 (98.41)  & 58.94 (101.04) & 99.52 (65.25)  \\
% RS                & 72.85 (101.01) & 24.90 (101.06) & 18.05 (101.28) &   \textcolor{red}{TO-DO}             \\
% MG                & 88.92 (101.20) & 74.06 (100.56) & 50.50 (100.83) & 21.63 (100.57) \\
% MG-MDNF           & 54.02 (101.41) & 29.20 (101.18) & 19.92 (101.48) & 19.25 (101.03) \\ \bottomrule
% \end{tabular}
\begin{tabular}{c|ccc|c}
\toprule
\multirow{2}{*}{Defense} & \multicolumn{3}{c|}{SNR-bounded PGD}                         & CW             \\ %\cline{2-5} 
                  & $20dB$       & $30dB$       & $40dB$       & $\epsilon=0.01$       \\ \midrule
None              & 102.5 & 91.4  & 58.9  & 106.3 (66.7)  \\
RS                & 72.9  & 24.9  & 18.1 &   43.8 (100.0)             \\
MG                & 88.9  & 74.1  & 50.5  & 93.8 (88.9) \\
MG-MDNF           & 54.0  & 29.2 & 19.9 & 56.3 (105.6) \\ \bottomrule
\end{tabular}
}
\end{table}

\begin{table}[tb]
    \centering
    \caption{\% WER~(lower is better) for \textbf{waveGAN defenses against untargeted PGD attacks}. 
    %WG by itself shows strong performance as a defense across all attacks, and the addition of noise flooding (WG-MDNF) produces little change in word error rate. 
    Results are averages over $100$ samples from the test clean split of the Librispeech dataset. Each column shows the results for a different threat model characterized by $\epsilon$.} 
    %All scenarios use a maximum iterations value of 100.}
    %Attack threat models are shown as $\epsilon$/max. iterations.}
    \label{tab:results_wavegan}
    \resizebox{0.48\textwidth}{!}{
    \begin{tabular}{c|c|ccc|cc}
\toprule
\multirow{2}{*}{Defense} & \multirow{2}{*}{Ben.} & \multicolumn{3}{c|}{PGD ($\ell_2$ norm)}     & \multicolumn{2}{c}{PGD ($\ell_\infty$ norm)} \\ %\cline{3-7} 
                  &                         & 1.5 & 1.0 & 0.5 & 0.01    & 0.005    \\ \midrule
None              & 12.7                   & 85.6   & 75.5   & 58.5   & 113.5      & 93.7        \\
WG                & 17.6                   & 39.2   & 33.3   & 26.2   & 41.7       & 29.7        \\
WG-MDNF           & 19.2                   & 40.5   & 33.3   & 26.6   & 41.0      & 29.5        \\ \bottomrule
\end{tabular}
}
    % \begin{tabular}{|p{1.8cm}|p{.8cm}|p{1.5cm}|p{1.5cm}|p{1.5cm}|p{1.5cm}|p{1.8cm}|}
    % \hline
    %  & Ben. & $\ell_2$/$1.5$/$100$ & $\ell_2$/$1.0$/$100$& $\ell_2$/$0.5$/$100$ &  $\ell_{\infty}$/$0.01$/$100$ & $\ell_{\infty}$/$0.005$/$100$ \\
    % \hline
    % \textbf{None}    & $12.73$ & $85.62$ & $75.52$ & $58.50$  & $113.46$ & $93.66$  \\
    % \hline
    % \textbf{WG}      & $17.59$ & $39.15$ & $33.30$ & $26.24$  & $41.71$ & $29.68$  \\
    % \hline
    % \textbf{WG-MDNF} & $19.19$ & $40.45$ & $33.30$ & $26.58$  & $40.99$  & $29.53$ \\
    % \hline
    % \end{tabular}
\end{table}
\section{Conclusion}
This paper introduces mel-domain noise flooding (MDNF) which leverages additive noise in the mel frequency domain followed by GAN-based re-synthesis of the time domain samples. While prior work (e.g. \cite{adversarial_attacks+defenses}) has shown that the mel conversion process can itself act as a defense, we find that not all generator models provide the same degree of robustness. We show that MDNF can substantially improve upon the defensive capability of a poorer performing generator, while having minimal impact on one with strong baseline robustness. Therefore noise flooding provides a quick and easy way to enhance defensive performance without the computational costs required to train a more robust generator from scratch. We demonstrate this result across a variety of untargeted and targeted PGD attacks as well as the targeted Calini-Wagner attack. We additionally find that MDNF is competitive with randomized smoothing, and substantially outperforms RS for stronger (i.e., lower SNR) attacks.

\pagebreak

\bibliographystyle{IEEEtran}

\bibliography{ref}

\end{document}